\documentclass[acmsmall,nonacm]{acmart}

\pdfoutput=1

\usepackage{booktabs}
\usepackage{subcaption}
\usepackage[frozencache]{minted}
\usepackage{listings}
\usepackage{bussproofs}
\usepackage{amsmath}
\usepackage{tikz}
\usepackage[nounderscore]{syntax}
\usetikzlibrary{arrows,positioning,calc}
\usepackage{mathrsfs}
\usetikzlibrary{ipe}
\graphicspath{{pics/}}
\usepackage{wrapfig}
\usepackage{caption}
\usepackage{mathtools}
\usepackage{needspace}

\newcommand{\codePlain}[1]{\texttt{#1}}
\newcommand{\coq}[1]{\lstinline{#1}}

\newcommand{\contract}{smart contract}

\newcommand{\contracts}{smart contracts}

\begin{document}

\title{Foundational Verification of Smart Contracts through Verified Compilation}
\author{Vilhelm Sjöberg}
\affiliation{
  \institution{CertiK}
  \country{USA}
}
\email{vilhelm.sjoberg@certik.io}

\author{Kinnari Dave}
\affiliation{
  \institution{CertiK}
  \country{USA}
}
\email{kinnari.dave@certik.io}

\author{Daniel Britten}
\affiliation{
  \institution{University of Waikato}
  \country{New Zealand}
}
\email{db130@students.waikato.ac.nz}

\author{Maria A Schett}
\affiliation{
  \institution{University College London}
  \country{England}
}
\email{mail@maria-a-schett.net}

\author{Xinyuan Sun}
\affiliation{
  \institution{CertiK}
  \country{USA}
}
\email{sxysun@certik.io}

\author{Qinshi Wang}
\affiliation{
  \institution{Princeton University}
  \country{USA}
}
\email{qinshiw@cs.princeton.edu}

\author{Sean Noble Anderson}
\affiliation{
  \institution{Portland State University}
  \country{USA}
}
\email{ander28@pdx.edu}

\author{Steve Reeves}
\affiliation{
  \institution{University of Waikato}
  \country{New Zealand}
}
\email{stever@waikato.ac.nz}

\author{Zhong Shao}
\affiliation{
  \institution{Yale University}
  \country{USA}
}
\email{zhong.shao@yale.edu}

\begin{abstract}
  Programs executed on a blockchain---\emph{smart contracts}---have
  high financial stakes; their correctness is crucial. We argue, that
  this correctness needs to be \emph{foundational}: correctness needs
  to be based on the operational semantics of their execution
  environment. In this work we present a foundational system---the
  \emph{DeepSEA system}---targeting the Ethereum blockchain as the
  largest smart contract platform. The DeepSEA system has a small but
  sufficiently rich programming language amenable for verification,
  the DeepSEA language, and a verified DeepSEA compiler.
  Together they enable true end-to-end verification for smart
  contracts.
  We demonstrate usability through two case studies: a realistic
  contract for Decentralized Finance and contract for crowdfunding.
\end{abstract}

\ccsdesc[500]{Software and its engineering~General programming languages}
\ccsdesc[300]{Social and professional topics~History of programming languages}

\keywords{Verification, Verified compilation, Blockchain, Smart Contracts}

\maketitle

\newcommand{\sna}[1]{{\color{purple} SNA: #1}}

\section{Introduction}
\emph{Blockchains} are distributed systems for recording and processing data. They are \emph{public}, i.e., not subject to permissions or  visibility-controls, \emph{decentralized}, i.e., not controlled by any single party, and use a consensus mechanism to ensure consistency and immutability of the widely and openly distributed data.
Blockchains were spearheaded by Bitcoin~\cite{nakamoto2008bitcoin} for simple transactions of passing bitcoins between accounts. The Ethereum blockchain generalized these transactions: users can execute arbitrary programs. These programs are compiled to bytecode for the \emph{Ethereum Virtual Machine} (EVM). This bytecode is then uploaded on the Ethereum blockchain and therefore immutable. Users can now call the bytecode of a program and such a transaction can change a range of state variables---also stored on the blockchain---by executing the Turing complete program. These programs are known as \emph{smart contracts}. Smart contracts programmatically transfer money following this immutable bytecode.
In the last few years, their promise and problems have become apparent. On the one hand Ethereum is widely used, in particular in \emph{decentralized finance} (DeFi). Cryptocurrencies trading has moved from centralized brokers to smart contracts and process a yearly trading volume of tens of billions of dollars. On the other hand these smart contracts have security flaws. Each year there are dozens of hacks resulting in losses of hundreds of thousands, or even millions of dollars.

Stakes as high as these attracted verification efforts. Many tools leverage automatic theorem proving by annotating smart contracts with pre- and post-conditions. These are translated into verification conditions (VCs) for an SMT solver. However, the theories that can be automatically handled by SMT solvers are limited. So while these tools worked well for the simple ``token'' contracts that dominated the landscape a few years ago, we believe more complex applications such as DeFi necessitate interactive proof assistants.
Even tools that can generate VCs for interactive proof (e.g. via intermediate languages like Why3) have weaknesses: the VC generation is not itself verified, so the verification tool itself could have correctness-critical bugs. This leads to the demand of \emph{foundational} systems, which are based on a formal semantics for the EVM machine itself~\cite{eth_isabelle, hildenbrandt2018kevm}. Proving correctness with respect to an operational semantics gives much higher confidence than proving VCs which were generated in an ad-hoc way. Some of the most critical and high-value contracts, e.g. the staking contract for Ethereum 2.0, have been foundationally verified~\cite{runtimeVerificationDeposit, runtimeVerificationDepositGitHub}. However, working directly on bytecode without the benefit of data abstraction is very laborious, which makes it hard to use these tools for large contracts.

In this work, we introduce the \emph{DeepSEA system}, which provides the best of both worlds. We define a \emph{programming language}, small but sufficiently rich to write realistic programs, and implement a \emph{verified compiler} for it. The compiler compiles the source program into
(1) EVM bytecode, and (2) a formal model which can be loaded into the Coq proof assistant to prove correctness properties.
The \emph{DeepSEA language} is designed to make programs amenable to verification. Programs are structured to enable modular verification, inspired by previous work on large program verification. It also tries to shield the user as much as possible from data representation concerns, by making the  Coq model use idealized ``mathematical'' data types like unbounded integers and finite maps, and separate out concerns about finite word-size and array bounds into a side-condition that can be verified before dealing with the main proof. We have tried the system on realistic smart contracts, and find it both: powerful and user-friendly.
The \emph{DeepSEA compiler} is implemented and verified in Coq, and the correctness theorem is internalized so one can check a theorem inside Coq that a particular smart contract specification is implemented by a given bytecode.

Writing a verified compiler for a blockchain language also revealed some interesting points that do not come up in more traditional verified compilers. The EVM tracks \emph{gas cost} and charges money for each step of execution and each byte stored on the blockchain. So the compiler correctness proof must take gas cost into account at each phase, and when phrasing the correctness theorem we must be careful to make sure it is still informative even if the compiler optimizes the gas cost. Gas costs for storing the bytecode on the blockchain also means that a slightly different set of optimizations are relevant: we want to minimize program size. Finally, unlike a compiler for C, we must be careful to distinguish undefined behavior which should be avoided by the programmer, e.g. an integer overflow, versus erroneous behavior that can always be triggered by an attacker, e.g. an out-of-gas exception.

Specifically we make the following contributions:

\begin{itemize}
\item an integrated system for proving smart contract correctness, consisting of a programming language (Section~\ref{sec:deepsea-pl}), tools for producing a Coq model of a contract (Section~\ref{sec:deepsea-usage}) and set of Coq tactics for reasoning about that model (Section~\ref{sec:invrunstateT}),
\item two case studies: a realistic DeFi contract (Section~\ref{sec:uniswap}) and a crowdfunding contract (Section~\ref{sec:crowdfunding}). We evaluate the proof effort of the crowdfunding contract by comparing it to a similar proof by a non-foundational tool, and find that DeepSEA is similarly user-friendly (Section~\ref{sec:crowdfunding-evaluation}).
\item a verified compiler from the DeepSEA language to EVM with
  several interesting features: we compile via a general-purpose
  intermediate language (Section~\ref{sec:minic}) with a novel memory
  model (Section~\ref{sec:memory-model}) and we use an extensible way to
  abstract away from datatype representation
  (Section~\ref{sec:datatype-representation}). Our correctness
  statement distinguishes datatype-related undefined behavior from
  genuinely unavoidable errors
  (Section~\ref{sec:correctness-statement}). The compiler correctness
  proofs track gas cost through all the phases
  (Section~\ref{sec:gas-handling}). We describe how to apply a large collection of
  automatically proven peephole optimization rules, which is
  particularly helpful in the EVM setting (Section~\ref{sec:optimizations}).
\end{itemize}

\section{The DeepSEA programming language}
\label{sec:deepsea-pl}

At the center of the DeepSEA system is a small programming
language. Each program is written as a set of objects with methods and
state variables, corresponding to contract methods and data which is
persistently stored on the blockchain.
We base this work on an earlier version of the DeepSEA
language~\cite{sjoberg:deepsea} which aims to support verified
operating system kernels and compiles into C and Coq. We now target
the EVM and Coq. As such, we inherit a set of language features to
support modular proofs. Each program is divided into a set of objects,
which are further grouped into ``layers'', and one can further use
``abstraction refinement'' to ascribe a simpler, manually written,
specification to part of a program. We have not used these features
extensively in our current blockchain case studies, but we expect them
to become relevant when verifying even larger smart contract applications.

DeepSEA supports primitive built-in datatypes like \codePlain{int} or
\codePlain{bool}, and allows users to define datatypes using the
keyword \codePlain{type}. It also supports arrays, hashmappings, and
\codePlain{structs} which are defined as a collection of
fields. However, complex data is currently only used for persistent
storage, while method arguments and local variables are word-sized.

Our compilation target is the Ethereum Virtual Machine (EVM). Each node in the Ethereum network has a bytecode interpreter which executes EVM bytecode as part of processing transactions. The EVM is \textit{quasi-Turing} complete: termination is guaranteed by executing code using a fuel called \textit{gas}. Each time you submit a transaction you pay for a certain number of units of gas, and the system can execute bytecode only as long as there is gas available to run it, with the gas cost for each instruction execution defined by the EVM specification~\cite{yellowpaper}.

The EVM is a stack based machine, which departs from the standard von Neumann architecture, in the way it handles storage of programs. Instead of storing the code of the program in a memory location which is generally accessible, it stores it in a virtual ROM-like memory, which can only be accessed by a special construct for EVM programs called the \textit{constructor}. The EVM model has three different types of regions to store data during program execution: \textit{stack}, \textit{memory} and \textit{storage}. The stack stores data which can be popped off or pushed on during program execution through special instructions designed exclusively for stack manipulation. The memory is a volatile word addressable byte array. The storage on the other hand is a persistent word addressable word array, which is a part of the VM state description. The standard implementation of hashmappings in Ethereum programming languages just writes directly to a storage address, which will be backed by a real hashtable in the bytecode interpreter.

\subsection{DeepSEA language by example: a token contract}

We now demonstrate the DeepSEA language by showing a small program: a
``token contract''. Tokens are fungible units of exchange, and the
contract maintains a database of how many tokens each user owns. In
effect, a token contract creates a new cryptocurrency which can be
transferred between users, and token contracts are by far the most
common type of contracts on Ethereum currently. In order to maintain
uniformity and aid ease of conversion, the Ethereum foundation has
created a standard known as ERC-20 which specifies what methods a
token contract shall implement.

The implementation of an ERC-20 compliant token in DeepSEA is laid out in the \codePlain{token.ds} file which we will describe in this section. It initially assigns a balance of 100000 token units to the
owner of the contract. The interface of this contract allows transfer of tokens, querying the balance of this particular token in an account and gives the constructor which initializes the owner's account
with a fixed supply of the tokens.
The entire DeepSEA program in \codePlain{token.ds} consists of a
single object and layer. Because the layer does not depend on any layer
below it is declared in the following manner:
\begin{minted}[fontsize=\footnotesize, linenos, breaklines]{coq}
layer TOK : [{}] TOKSig = {
  tok = Tok
} assert "Invariant.inv"
\end{minted}

Here, the square brackets followed by the layer declaration enclose
layer dependencies---in our case the empty list of signatures
\codePlain{\{\}} indicates a low-level layer with no dependencies
beneath.

\textit{Invariants} for objects (or layers) are declared with
\codePlain{assert}. The user must prove that invariants are preserved
by all methods declared inside objects. The DeepSEA compiler generates
a set of verification side-conditions, and the invariant is also
available when proving those.

In our example, the \codePlain{Invariant.inv} condition requires that the balance of the token being implemented in all accounts is non-negative and that the sum of the balances must add up to the total supply of the token.
Finally, \codePlain{tok} is the object enclosed within this layer. Its signature is given as follows:
\begin{minted}[fontsize=\footnotesize, linenos, breaklines]{coq}
object signature ERC20Interface = {
  constructor : unit -> unit;                    transfer : address * int -> bool;
  const totalSupply : unit -> int;               approve : address * int -> bool;
  const balanceOf : address -> int;              transferFrom : address * address * int -> bool
}
\end{minted}

The \codePlain{Tok} object is an instance of an object with the above signature and highlights various important features of DeepSEA:

\begin{minted}[fontsize=\footnotesize, linenos, breaklines]{coq}
object Tok : ERC20Interface {
  let balances : mapping[address] int := mapping_init
  let allowances : mapping[address] mapping[address] int := mapping_init
  let constructor () = balances[msg_sender] := 100000
  let balanceOf tokenOwner = let bal = balances[tokenOwner] in bal
  let transfer(toA, tokens) =
    let fromA = msg_sender in let from_bal = balances[fromA] in let to_bal = balances[toA] in
    assert (fromA <> toA /\ from_bal >= tokens);
    balances[fromA] := from_bal-tokens;
    balances[toA] := to_bal+tokens;
    true
    ...
  }
\end{minted}

We can see that DeepSEA does not use dynamic memory allocation, and  the \codePlain{balances} and \codePlain{allowances} mappings are initialized using the placeholder initializer
\codePlain{mapping\_init} which initializes them with zeros (a no-op on the EVM).

\subsection{Verifying smart contracts}
\label{sec:deepsea-usage}

Now we discuss the Coq specifications which DeepSEA generates, and the choice of datatype representation in these specifications.
The DeepSEA compiler dsc generates a directory \codePlain{token} as a part of the compilation process of the DeepSEA source file \codePlain{token.ds}. As a result the \codePlain{token} directory is populated with
the files \codePlain{LayerTOK.v} and \codePlain{ObjTokCodeProofs.v} among others.
The file \codePlain{LayerTOK.v} is where the Coq specifications of the methods in various objects defined in the layer are written. Since the \codePlain{token.ds} contract consists of exactly one
layer \codePlain{TOK} and one object \codePlain{Tok} inside, with no dependency on any underlying layer, it is considered to be in kernel mode.
The data types used to represent the variables \codePlain{balances} and \codePlain{allowances} are declared as fields of the record \codePlain{global\_abstract\_data\_type}:
\begin{minted}[fontsize=\footnotesize, linenos, breaklines]{coq}
Record global_abstract_data_type : Type := Build_global_abstract_data_type
  { Tok_balances : Int256Tree.t Z32;
    Tok_allowances : Int256Tree.t (Int256Tree.t Z32) }
\end{minted}
Every Gallina\footnote{Gallina is the specification language of Coq.} specification generated for the methods defined in this object uses this data representation in its definition, as is evident from the \codePlain{Tok\_transfer\_opt} definition below.
Since the program is defined as a single layer, and is not a vertical composition of two layers, the instance generated for the \codePlain{\_overlay\_spec} is the empty record and the 
\codePlain{\_underlay\_spec} is simply the instance \codePlain{GlobalLayerSpec} declared for the entire program. Similarly, the object variables \codePlain{balances, allowances} are represented by the 
definitions ending with \codePlain{\_var} as defined in the \codePlain{LayerTOK.v}. The \codePlain{constructor} is represented by the definition ending with \codePlain{\_constructor}.

We now illustrate the specification generation mechanism of DeepSEA and the proof generation of its equivalence to the code through its treatment of the \codePlain{transfer} function defined as a part of the object definition above.
The generated \coq{Tok_transfer} definition has type \codePlain{function\_constr} which is a record type parametrized by an element of type \codePlain{LayerSpecClass}. The Gallina specification is the definition \codePlain{Tok\_transfer\_opt} which is of type  \codePlain{int256 -> Z32 ->  machine\_env GetHighData -> DS bool}.
Also generated by dsc are automatic proofs of the fact that the ASTs that it generates for all the methods are well-typed:
\begin{minted}[fontsize=\footnotesize, linenos, breaklines]{coq}
Lemma Tok_transfer_wf : synth_func_wellformed Tok_transfer.
  Proof. solve_wellformed. Defined.
\end{minted}
The dsc generated specification for the transfer method is given as follows. Note that \coq{address} is a type alias for \coq{int256}.

\begin{minipage}[t]{0.55\linewidth}
\begin{minted}[fontsize=\footnotesize, linenos, breaklines]{coq}
Tok_transfer_opt memModelOps f f0 me : DS bool :=
spec1 <- ret (me_caller me);;
spec2 <- gets
  (fun s  => get_default 0 f (Tok_balances s));;
(v <- ret (negb (eq spec1 f) && (spec2 >=? f0));;
  guard v);;
modify
  (fun s => update_Tok_balances
  (set spec1 (spec2 - f0) (Tok_balances s)) s);;
modify
  (fun s => update_Tok_balances
  (set f (spec3 + f0) (Tok_balances s)) s);;
  ret true.
\end{minted}
\end{minipage}
\begin{minipage}[t]{0.45\linewidth}
\begin{minted}[fontsize=\footnotesize, linenos=false, breaklines]{coq}
(* Corresponding .ds file: *)  
(* let fromA = msg_sender *)
(* let from_bal = balances[fromA] *)

(* assert (fromA<>toA              *)
(*        /\ from_bal>=tokens)     *)

        
(* balances[fromA] := from_bal-tokens *)


(* balances[toA] := to_bal+tokens    *)
\end{minted}
\end{minipage}

The \codePlain{\_opt} suffix in the name of the generated
specification refers to the use of the \textit{option monad} to capture the effect of assertion failures.
The effects of commands are captured in the specification using the monadic type:
\begin{minted}[fontsize=\footnotesize, breaklines]{coq}
Definition DS := stateT GetHighData option. (* DS a = GetHighData -> option(a*GetHighData) *)
\end{minted}

Since the specification is what the user will reason about in the
proofs, we want it to correspond in an obvious way to the input
program, and it mostly does except apart from being written in Coq
syntax instead of DeepSEA syntax and using monadic combinators for the
imperative operations.

If we look at the details, in some ways the Coq representation is
actually more high-level and easier to reason about than the input
program. While compiled Ethereum hash tables do not keep track of
which keys are stored in them, in the Coq version we represent the
balances table with a Int256Tree data type which does remember its
keys. The executable code still cannot make use of the keys, but
theorems about the code can talk about them. Similarly, while the
executable program uses bounded 256-bit integers, the representation
uses unbounded mathematical integers so that theorems about ordinary
arithmetic applies. Instead, the DeepSEA system generates a set of
separate side conditions for the user to prove there is no overflow.

We can then write theorems about the generated specifications just as
we would for any monadic Coq function.
Consider, for example this lemma about the \codePlain{tok\_transfer\_opt} satisfying the invariant that the sum of the balances of the token in all accounts remains constant after a transfer:
\begin{minted}[fontsize=\footnotesize, linenos, breaklines]{coq}
Theorem transfer_constant_balances_sum : forall toA n d d' me b,
runStateT (Tok_transfer_opt toA n me) d = Some (b, d') -> balances_sum d' = balances_sum d.
\end{minted}

Let us consider the above theorem statement. Note that the variables characterizing the data needed for the contract to execute any of its methods are the maps \codePlain{balances} and \codePlain{allowances}. Hence,
the state in this case is represented by \codePlain{global\_abstract\_data\_type} as defined above. The monadic evaluator \codePlain{runStateT} evaluates the effect of calling the transfer method in state 
$d$ and returns a value of type \codePlain{DS bool} which is an option on the \codePlain{bool * state} pair. In case of a successful run (i.e no assertion failure was reported) the theorem states that the sum of the balances
in the new state $d'$ is the same as that in the state $d$. DeepSEA defines a few custom tactics to help proving such results by simplifying monadic computations in the context.

\section{Case studies}
\label{sec:case-studies}

The small token contract discussed above is easy to verify using many
methods. In this section we discuss two larger case studies, where the
power of reasoning about high-level semantics in an interactive proof
assistant is crucial.

\subsection{Uniswap-style market maker}
\label{sec:uniswap}
 An example that highlights the importance of having a language which enables reasoning about code at ``all levels'' is the automated market maker smart contract prototype.
 Constant-product automated market makers is one of the most important and popular
 smart contract types in decentralized finance (DeFi). They support exchange of tokens on blockchain-based platforms without a centralized party keeping track of such transactions. This is accomplished
by maintaining liquidity reserves of two tokens, and determining the exchange rate using a mathematical formula which is a function of the token reserves. Various
users can provide liquidity of tokens in return for receiving the
transaction fees charged by the contract to the traders.

These contracts also serve as oracles which can be queried for the
current market rate for particular exchanges. Other contracts,
e.g. for lending and margin trading, use this price information to
value token deposits used as collaterals.
Hence, manipulating such oracles can prove to be very beneficial to hackers. 

\subsubsection{Contract implementation}

The smart contract written in DeepSEA uses the Uniswap v2 protocol as a blueprint. In the DeepSEA setup, the entire contract is defined as a layer
AMM on top of an underlay layer called the AMMLIB. The AMMLIB layer consists of three objects: two ERC20 tokens which are to be swapped and a liquidity token. The AMM layer acts as the interface for the contract. This layer consists of an object of type AMMInterface, which defines the methods that provide all the functionalities of the protocol. The methods in this object signature are given as follows: \\
\begin{itemize}
    \item \coq{simpleSwap0} allows the transfer of one token to the contract to be exchanged for the other, and returns the amount of the second token to be received in return;
    \item \coq{mint} allows the transfer of liquidity to a liquidity pool for a liquidity provider;
    \item \coq{burn} allows a liquidity provider to withdraw liquidity from a pool;
    \item \coq{sync} is a recovery mechanism method to prevent the market for the given pair from being stuck in case of low reserves;
    \item \coq{skim}  prevents any user from depositing more tokens in any reserve than the maximum limit, to prevent overflow;
    \item \coq{k} tracks the product of the reserves;
    \item \coq{quote0} returns the equivalent amount of the second token, given an amount of the first token and current reserves in the contract;
    \item \coq{getAmountOut0} returns the maximum possible amount of a token than can be gained in exchange for a particular input amount of the other token and that of the reserves;
    \item \coq{getAmountIn0} returns the amount of a given token that must be input in order to obtain the desired amount of the other token under the given reserves.
    
\end{itemize}
Compared to the Uniswap protocol, we have made a few simplifications. Unlike Uniswap, which offers the option of switching on/off the protocol fee, the DeepSEA contract does not model protocol fees. Moroever, instead of using the above mentioned square root formula to calculate the share of minted liquidity tokens for a liquidity provider, the DeepSEA contract uses the product and burns the first 1000 coins, as in Uniswap v2. The price oracle mechanism is based on Uniswap v1, and the DeepSEA contract does not support flash swaps. In order to make our contract completely ABI-compatible with the original we may add these features. However, the DeepSEA AMM contract already offers all the core functionality offered by the Uniswap protocol, and it contains everything that is relevant to the specification that we are verifying. As such, our proof is an example of verifying a realistic contract.

The exchange of two tokens is enabled by the simple swap function declared as a part of the AMMInterface signature above. The method definition is given as follows:
\begin{minted}[fontsize=\footnotesize, linenos, breaklines]{coq}
let simpleSwap0 (toA) =
    let reserve0 = _reserve0 in
    let reserve1 = _reserve1 in
    let balance0 = erc20Token0.balanceOf(this_address) in
    let balance1 = erc20Token1.balanceOf(this_address) in
    let amount0In = balance0 - reserve0 in
    let token0 = _token0 in
    let token1 = _token1 in
    assert (toA <> token0 /\ toA <> token1);
    assert (amount0In > 0);
    assert (reserve0 > 0 /\ reserve1 > 0);
    let amountInWithFee = amount0In * 997 in
    let numerator = amountInWithFee * reserve1 in
    let denominator = reserve0 * 1000 + amountInWithFee in
    let result = numerator / denominator in
    let success = erc20Token1.transfer(toA, result) in
    assert (success);
    let balance0 = erc20Token0.balanceOf(this_address) in
    let balance1 = erc20Token1.balanceOf(this_address) in
    _reserve0 := balance0;
    _reserve1 := balance1;
    let resultU = result in
    resultU	
\end{minted}

The main object of the contract is 144 lines, to be compared with 200
lines of the original \codePlain{UniswapV2Pair.sol} contract which we
are imitating. Although these are not very large by conventional
software development terms, it is big enough to show practical issues
with the development system, for example we had to add a live-variable
analysis phase to the DeepSEA compiler to fit all the local variables
on the EVM stack.

\subsubsection{Proof}

Recent work~\cite{angeris2019analysis} develop a mathematical analysis
of this style of AMMs. One of their theorems is was a lower bound on
the cost of trades to move the reported price of the contract by a
certain amount. This theorem provides  measures of the various parameters that needed to be controlled in order to prevent oracle-manipulation attacks.

Using DeepSEA we formally proved this theorem in the Coq, connecting
the parameters involved directly to the generated bytecode, while
being rigorous about integers versus real numbers.  We state and prove the \codePlain{cost\_of\_manipulation\_min} result under the assumption that a swap operation was
successful (as indicated by the hypothesis \codePlain{del\_alp}):
\begin{minted}[fontsize=\footnotesize, linenos, breaklines]{coq}
Hypothesis del_alp : runStateT (AutomatedMarketMaker_simplSwap0_opt
      toA (make_machine_env a)) s = Some (r , s').

Theorem cost_of_manipulation_min : eps >= 0 ->
      cost_of_manipulation_val >= (IZR (reserve_beta s)) * (5/100) * sqrt (eps) \/
      cost_of_manipulation_val >= (IZR (reserve_beta s)) * (1/48) * (eps^2).
\end{minted}

The DeepSEA system generated specification of the
\codePlain{simpleSwap0} method enables us to establish a lower bound
for this cost and connect it directly to the bytecode generated for
this contract. The details of the proof have been published as a
workshop paper~\cite{amm_paper}. Here, we just note that it requires
some nontrivial mathematical results, e.g. the the Taylor series
approximation  twice differentiable functions. We rely on existing
libraries like Coq-interval~\cite{martin2013certified} and
Coquelicot~\cite{boldo2015coquelicot} for these. This highlights the
benefit of using an interactive theorem prover which can handle
arbitrary mathematical reasoning.

\subsection{Crowdfunding}
\label{sec:crowdfunding}
The automated market maker example shows that DeepSEA can verify a complex \contract{}. We next consider the usability when writing interactive proofs in DeepSEA. Two other Coq-based systems for reasoning about \contracts{},  Scilla~\cite{sergey2019safer} and ConCert~\cite{annenkov2020concert}, both use a Crowdfunding contract to evaluate their system. In this section we use the same benchmark, and consider a DeepSEA Crowdfunding \contract{} based upon the Scilla version~\cite{scilla_coq_repo}. The correctness proof is roughly as complex as for the Scilla version. Scilla and ConCert do not have verified backends and define the Coq representation of the contract in a way which is convenient to reason about, so we consider this a sign that DeepSEA's representation is adequate.

The Crowdfunding \contract{} decentralises fundraising for a project by setting a funding goal and time limit, and letting users make donations. If the goal is met within the time limit, the project receives the funds and if not, then all backers can reclaim their funds. Listing \ref{list:donateCrowdfunding} shows the \coq{donate} function in the DeepSEA version of the Crowdfunding contract.

\begin{listing}[h]
\begin{minted}[fontsize=\footnotesize, linenos, breaklines]{ocaml}
let donate () =
  assert(msg_sender <> this_address);
  assert(msg_value >= 0);
  let bs = backers in
  let blk = block_number in
  let _max_block = max_block in
  if (blk > _max_block) then
    begin
      emit Message(_deadlinePassed_msg);
      assert(false) (* Revert: do not accept funds *)
    end
  else
    begin
      let backed_amount = backers[msg_sender] in
      if (backed_amount = 0) then
          backers[msg_sender] := msg_value
      else
        begin
          emit Message(_alreadyDonated_msg);
          assert(false) (* Revert: do not accept funds *)
        end
    end
\end{minted}
\caption{The \coq{donate} function of the Crowdfunding smart contract in DeepSEA}
  \label{list:donateCrowdfunding}
\end{listing}

\subsubsection{Reasoning about monadic programs in Coq}
\label{sec:invrunstateT}

One difference between our proofs and the Scilla proofs is that our Coq model of the contract is specified as a function in the state monad rather than an inductively defined relation. This makes the connection between the DeepSEA source program and the Coq model more obvious, and one can extract code from specifications which could be helpful for QuickChick-style automated testing~\cite{denes2014quickchick}. From a practical perspective, this means we need some changes to the proof tactics we use, particularly \codePlain{inversion}.

If the semantics of a programming language is given by an inductively defined step relation, then the \codePlain{inversion} tactic can take e.g. an assumption about a sequenced command \codePlain{steps st1 (c1;;c2) st2} and conclude that there exists some intermediate state \codePlain{st} such that \codePlain{steps st1 c1 st} and \codePlain{steps st c2 st2}. Similarly, if the step rules specify how the state is updated, the tactic introduces equalities into the context. Contract correctness theorems have the form ``if a state is reachable by a sequence of steps, then \dots'', so their proofs heavily rely on using inversion to decompose the sequence.

In our case, the semantics of code sequencing \codePlain{c1;;c2} is specified by the \codePlain{bind} function for the state monad, so the built-in inversion tactic does not do anything, and instead we prove lemmas saying e.g. ``if \codePlain{runStateT (c1;;c2) st1 = Some (return\_value, st2)} then there exists an intermediate state \codePlain{st} such that \dots'', and similarly for equations, and then provide a custom-written tactic \codePlain{inv\_runStateT} to apply them.

While this is simple in theory, we found that it required some care in practice because some of Coq's built-in methods take more than linear time in the size of the terms they work on. For small examples there are no issues, but for the large method specifications from e.g. the market maker example, they get intolerably slow. For the \codePlain{subst} tactic, which exploits equalities, we needed to turn off Coq's \codePlain{Regular Subst Tactic} flag. And the \codePlain{destruct} tactic, which uses ``there exists'' assumptions, seems to take at least quadratic time, so we can not use a literal existential statement in the above lemma. Instead we use a Church-encoded version.

With \codePlain{inv\_runStateT}, reasoning about inversion of monadic functions is just as convenient as inversion of inductive relations. We also define the tactic \coq{inv_runStateT_branching} which also branches at every conditional statement.

\subsubsection{Proofs}

The correctness property we aimed to prove was that if a backer made a donation but the goal was not met, then the backer is able to get their donation back. As first discussed in~\cite{sergey_scilla_2018}, the following three properties taken together provide this guarantee.
  \begin{enumerate}
    \item
      \emph{The contract's balance is not less than the recorded sum of all the backers' contributions.}
    \item 
      \emph{If the goal is not met and the time expires, it is possible for backers to withdraw their funds.}
    \item 
      \emph{The recorded balances are not altered except when depositing and valid withdrawals.}
  \end{enumerate}

DeepSEA does not include a model of blockchain interactions, i.e. what sequences of method calls are possible. Currently, the user must manually define that as part of the statement of the correctness theorem. In the case of the Crowdfunding contract we define a \coq{step} relation which makes method calls while keeping track of blockchain parameters like user balances and block numbers. We then define a reachability predicate \coq{ReachableFromBy} in terms of the step relation. This predicate describes which states are reachable from a starting state by a list of intermediate steps and actions.

\begin{lemma}
  \label{lem:sufficient-funds-safe}
  The contract's balance is not less than the recorded sum of all the backers' contributions.
\end{lemma}

As part of the user defined model of blockchain interactions the predicate \coq{Safe} is defined, which is \coq{True} when the given predicate defined over an individual state of the blockchain model is true for all reachable states. This is useful for defining the lemma. We can state that the contract's balance is not less than the recorded sum of all the backers' contributions in Coq as follows:

\begin{minted}[fontsize=\footnotesize, linenos, breaklines]{coq}
Definition balance_backed state := (Crowdfunding_funded (contract_state state)) = false
  -> sum (Crowdfunding_backers (contract_state state)) <= (balance state (contract_address))
     /\ (forall k v,
           get k (Crowdfunding_backers (contract_state state)) = Some v
           -> ((v >= 0) /\ (v < Int256.modulus))).
Lemma sufficient_funds_safe : Safe balance_backed.
\end{minted}

The proof uses induction on \coq{ReachableFromBy}. After proving that the initial state satisfies the invariant \coq{balance_backed}, each of the cases representing different possible blockchain actions are considered.

When reasoning about the \coq{donate} function, lemmas about the datatype used for the backers mapping are required. At one point a lemma stating that if an item $k$ in a mapping initially has the value zero, then setting $k$ to the value $v$ means the sum now is the original sum plus $v$. The proof heavily uses the \coq{inv_runStateT_branching} tactic mentioned earlier. While completing these proofs it also became clear how useful it is that DeepSEA enables a quick workflow for switching between proving and improving the contract's implementation (after finding the occasional bug in the contract during the proof process).

\begin{lemma}
  \label{lem:can-claim-back}
  If the goal is not met and the time expires, it is possible for backers to withdraw their funds.
\end{lemma}

This lemma is proved in a similar manner, with the  exception that it is also necessary here to prove that the contract \emph{does not revert} in certain circumstances, which helps guarantee that it is possible for backers to withdraw their funds. The proof makes use of the previous lemma in order to guarantee that the \contract{} has sufficient funds to refund the backer. This distinguishes our proof from the Scilla version which is not structured in a way which makes this link between the lemmas in Coq.

\begin{lemma}
  \label{lem:donation-preserved}
  The recorded balances are not altered except when depositing and valid withdrawals.
\end{lemma}

The final lemma is stated using a temporal predicate, claiming the following: in all states since a donation is recorded, that donation remains recorded as long as all intermediate states do not involve a claim from the donor. In other words, a donation is recorded since that donation was recorded, as long as there were no claims from the donor in the meantime. This lemma is expressed formally in Coq as shown here.

\begin{minted}[fontsize=\footnotesize, linenos, breaklines]{coq}
Lemma donation_preserved : forall (a : addr) (d : Z),
  (donation_recorded a d) `since` (donation_recorded a d) `as-long-as` (no_claims_from a).
\end{minted}

The ability to state lemmas such as this, which involve reachability and properties about traces of smart contract execution, highlights the expressivity that is a strength of DeepSEA coupled with the user-defined blockchain model used for the Crowdfunding contract.

The proof proceeds similarly to Lemma \ref{lem:sufficient-funds-safe}, by induction on the reachability predicate. The inductive hypothesis as well as the assumption about the intermediate states having no claim from the donor is sufficient to establish that the second to last state of the trace has the donation recorded. The proof then shows that this final step also preserves the record of the donation.

\subsubsection{Proofs of side conditions}
As mentioned earlier, for a DeepSEA proof to be valid it is necessary to also prove certain automatically generated goals, for example relating to the absence of overflows. In the case of the Crowdfunding contract all the side conditions can be trivially discharged.

\subsubsection{Evaluation}
\label{sec:crowdfunding-evaluation}
The proof of Lemma \ref{lem:sufficient-funds-safe} uses about 260 Coq tactic invocations. The equivalent proof in Scilla uses about 100 (though is written in the terser mathcomp proof style). In ConCert, it is about 55. The proofs all follow a similar overall structure.

\ 

These proofs about the Crowdfunding smart contract and the associated user-defined blockchain model are available at: \url{https://github.com/Coda-Coda/Crowdfunding/tree/foundational-verification-paper}.

\section{Verified compilation} \label{sec:compiler}

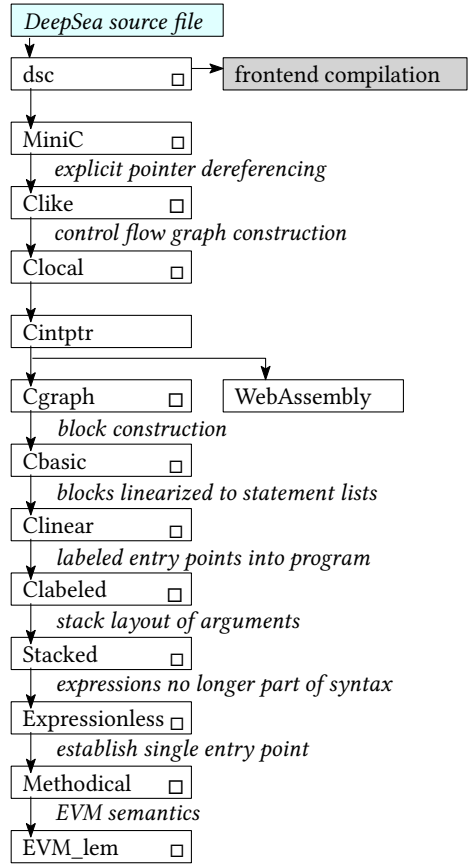
\begin{wrapfigure}[29]{r}{0.45\linewidth}
  \centering
  \tikzstyle{ipe stylesheet} = [
  ipe import,
  even odd rule,
  line join=round,
  line cap=butt,
  ipe pen normal/.style={line width=0.4},
  ipe pen fat/.style={line width=1.2},
  ipe pen heavier/.style={line width=0.8},
  ipe pen ultrafat/.style={line width=2},
  ipe pen normal,
  ipe mark normal/.style={ipe mark scale=3},
  ipe mark large/.style={ipe mark scale=5},
  ipe mark small/.style={ipe mark scale=2},
  ipe mark tiny/.style={ipe mark scale=1.1},
  ipe mark normal,
  /pgf/arrow keys/.cd,
  ipe arrow normal/.style={scale=7},
  ipe arrow large/.style={scale=10},
  ipe arrow small/.style={scale=5},
  ipe arrow tiny/.style={scale=3},
  ipe arrow normal,
  /tikz/.cd,
  ipe arrows, 
  <->/.tip = ipe normal,
  ipe dash normal/.style={dash pattern=},
  ipe dash dotted/.style={dash pattern=on 1bp off 3bp},
  ipe dash dash dot dotted/.style={dash pattern=on 4bp off 2bp on 1bp off 2bp on 1bp off 2bp},
  ipe dash dash dotted/.style={dash pattern=on 4bp off 2bp on 1bp off 2bp},
  ipe dash dashed/.style={dash pattern=on 4bp off 4bp},
  ipe dash normal,
  ipe node/.append style={font=\normalsize},
  ipe stretch normal/.style={ipe node stretch=1},
  ipe stretch normal,
  ipe opacity 10/.style={opacity=0.1},
  ipe opacity 30/.style={opacity=0.3},
  ipe opacity 50/.style={opacity=0.5},
  ipe opacity 75/.style={opacity=0.75},
  ipe opacity opaque/.style={opacity=1},
  ipe opacity opaque,
]
\definecolor{red}{rgb}{1,0,0}
\definecolor{blue}{rgb}{0,0,1}
\definecolor{brown}{rgb}{0.647,0.165,0.165}
\definecolor{darkblue}{rgb}{0,0,0.545}
\definecolor{darkcyan}{rgb}{0,0.545,0.545}
\definecolor{darkgray}{rgb}{0.663,0.663,0.663}
\definecolor{darkgreen}{rgb}{0,0.392,0}
\definecolor{darkmagenta}{rgb}{0.545,0,0.545}
\definecolor{darkorange}{rgb}{1,0.549,0}
\definecolor{darkred}{rgb}{0.545,0,0}
\definecolor{gold}{rgb}{1,0.843,0}
\definecolor{gray}{rgb}{0.745,0.745,0.745}
\definecolor{green}{rgb}{0,1,0}
\definecolor{lightblue}{rgb}{0.678,0.847,0.902}
\definecolor{lightcyan}{rgb}{0.878,1,1}
\definecolor{lightgray}{rgb}{0.827,0.827,0.827}
\definecolor{lightgreen}{rgb}{0.565,0.933,0.565}
\definecolor{lightyellow}{rgb}{1,1,0.878}
\definecolor{navy}{rgb}{0,0,0.502}
\definecolor{orange}{rgb}{1,0.647,0}
\definecolor{pink}{rgb}{1,0.753,0.796}
\definecolor{purple}{rgb}{0.627,0.125,0.941}
\definecolor{seagreen}{rgb}{0.18,0.545,0.341}
\definecolor{turquoise}{rgb}{0.251,0.878,0.816}
\definecolor{violet}{rgb}{0.933,0.51,0.933}
\definecolor{yellow}{rgb}{1,1,0}
\definecolor{black}{rgb}{0,0,0}
\definecolor{white}{rgb}{1,1,1}
\begin{tikzpicture}[ipe stylesheet]
  \filldraw[fill=lightgray]
    (144, 788) rectangle (236, 776);
  \node[ipe node, font=\small]
     at (68.378, 538.866) {Expressionless};
  \draw[shift={(64, 548)}, yscale=0.75]
    (0, 0) rectangle (68, -16);
  \node[ipe node, font=\small]
     at (68.378, 562.866) {Stacked};
  \draw[shift={(64, 572)}, yscale=0.75]
    (0, 0) rectangle (68, -16);
  \draw[-{ipe pointed[ipe arrow small]}]
    (71.4068, 560.717)
     -- (71.4068, 545.602);
  \node[ipe node, font=\small]
     at (68.378, 586.866) {Clabeled};
  \draw[shift={(64, 596)}, yscale=0.75]
    (0, 0) rectangle (68, -16);
  \draw[-{ipe pointed[ipe arrow small]}]
    (71.4068, 584.717)
     -- (71.4068, 569.602);
  \node[ipe node, font=\small]
     at (68.378, 610.866) {Clinear};
  \draw[shift={(64, 620)}, yscale=0.75]
    (0, 0) rectangle (68, -16);
  \draw[-{ipe pointed[ipe arrow small]}]
    (71.4068, 608.717)
     -- (71.4068, 593.602);
  \node[ipe node, font=\small]
     at (68.378, 634.866) {Cbasic};
  \draw[shift={(64, 644)}, yscale=0.75]
    (0, 0) rectangle (68, -16);
  \draw[-{ipe pointed[ipe arrow small]}]
    (71.4068, 632.717)
     -- (71.4068, 617.602);
  \node[ipe node, font=\small]
     at (68.378, 658.866) {Cgraph};
  \draw[shift={(64, 668)}, yscale=0.75]
    (0, 0) rectangle (68, -16);
  \draw[-{ipe pointed[ipe arrow small]}]
    (71.4068, 656.717)
     -- (71.4068, 641.602);
  \node[ipe node, font=\small]
     at (68.378, 682.866) {Cintptr};
  \draw[shift={(64, 692)}, yscale=0.75]
    (0, 0) rectangle (68, -16);
  \draw[-{ipe pointed[ipe arrow small]}]
    (71.4068, 680.717)
     -- (71.4068, 665.602);
  \node[ipe node, font=\small]
     at (68.378, 490.866) {EVM\_lem};
  \draw[shift={(64, 500)}, yscale=0.75]
    (0, 0) rectangle (68, -16);
  \node[ipe node, font=\small]
     at (68.378, 514.866) {Methodical};
  \draw[shift={(64, 524)}, yscale=0.75]
    (0, 0) rectangle (68, -16);
  \draw[-{ipe pointed[ipe arrow small]}]
    (71.4068, 512.717)
     -- (71.4068, 497.602);
  \draw[-{ipe pointed[ipe arrow small]}]
    (71.4068, 536.717)
     -- (71.4068, 521.602);
  \node[ipe node, font=\small]
     at (68.378, 706.866) {Clocal};
  \draw[shift={(64, 716)}, yscale=0.75]
    (0, 0) rectangle (68, -16);
  \draw[-{ipe pointed[ipe arrow small]}]
    (71.4068, 704.717)
     -- (71.4068, 689.602);
  \node[ipe node, font=\small]
     at (68.378, 730.866) {Clike};
  \draw[shift={(64, 740)}, yscale=0.75]
    (0, 0) rectangle (68, -16);
  \draw[-{ipe pointed[ipe arrow small]}]
    (71.4068, 728.717)
     -- (71.4068, 713.602);
  \node[ipe node, font=\small]
     at (68.378, 754.866) {MiniC};
  \draw[shift={(64, 764)}, yscale=0.75]
    (0, 0) rectangle (68, -16);
  \draw[-{ipe pointed[ipe arrow small]}]
    (71.4068, 752.717)
     -- (71.4068, 737.602);
  \node[ipe node, font=\small]
     at (80, 720) {\emph{control flow graph construction}};
  \filldraw[shift={(64, 808)}, xscale=5, yscale=0.75, fill=lightcyan]
    (0, 0) rectangle (16, -16);
  \node[ipe node, font=\small]
     at (68.756, 798.866) {\emph{DeepSea source file}};
  \node[ipe node, font=\small]
     at (68.378, 778.866) {dsc};
  \draw[shift={(64, 788)}, yscale=0.75]
    (0, 0) rectangle (68, -16);
  \draw[-{ipe pointed[ipe arrow small]}]
    (71.4068, 776.717)
     -- (71.4068, 761.602);
  \draw[-{ipe pointed[ipe arrow small]}]
    (72, 676)
     -- (160, 676)
     -- (160, 668);
  \node[ipe node, font=\small]
     at (148.378, 658.866) {WebAssembly};
  \draw[shift={(144, 668)}, yscale=0.75]
    (0, 0) rectangle (68, -16);
  \node[ipe node, font=\small]
     at (148.378, 778.866) {frontend compilation};
  \draw[-{ipe pointed[ipe arrow small]}]
    (132, 784)
     -- (144, 784);
  \draw[-{ipe pointed[ipe arrow small]}]
    (70.9372, 795.393)
     -- (70.9373, 788.053);
  \node[ipe node, font=\small]
     at (81.134, 647.072) {\emph{block construction}};
  \node[ipe node, font=\small]
     at (80.756, 599.083) {\emph{labeled entry points into program}};
  \node[ipe node, font=\small]
     at (80.756, 575.461) {\emph{stack layout of arguments}};
  \node[ipe node, font=\small]
     at (80.756, 551.85) {\emph{expressions no longer part of syntax}};
  \node[ipe node, font=\small]
     at (124.06, 777.634) {$\Box$};
  \node[ipe node, font=\small]
     at (124.06, 753.828) {$\Box$};
  \node[ipe node]
     at (123.304, 730.023) {$\Box$};
  \node[ipe node, font=\small]
     at (123.682, 705.462) {$\Box$};
  \node[ipe node]
     at (122.926, 657.851) {$\Box$};
  \node[ipe node, font=\small]
     at (123.304, 633.29) {$\Box$};
  \node[ipe node, font=\small]
     at (123.304, 609.106) {$\Box$};
  \node[ipe node]
     at (122.548, 585.3) {$\Box$};
  \node[ipe node, font=\small]
     at (123.682, 561.495) {$\Box$};
  \node[ipe node, font=\small]
     at (123.682, 537.312) {$\Box$};
  \node[ipe node]
     at (122.926, 513.506) {$\Box$};
  \node[ipe node, font=\small]
     at (123.682, 490.079) {$\Box$};
  \node[ipe node, font=\small]
     at (80, 744) {\emph{explicit pointer dereferencing}};
  \node[ipe node, font=\small]
     at (80.756, 623.083) {\emph{blocks linearized to statement lists}};
  \node[ipe node, font=\small]
     at (80.756, 527.85) {\emph{establish single entry point}};
  \node[ipe node, font=\small]
     at (80.756, 503.85) {\emph{EVM semantics}};
\end{tikzpicture}
  \caption{Phases of the Compiler.}
  \label{fig:compiler-overview}
\end{wrapfigure}

The DeepSEA compiler consists of an unverified frontend called dsc and a verified compiler. The frontend dsc is implemented in OCaml. It typechecks the DeepSEA source code and also generates a directory of Coq files containing the AST of the program and proofs connecting each layer declared in the source code. The generated Coq files then import the verified compiler, which translates the AST into EVM bytecode via an intermediate language called MiniC.

The dsc part is 11,000 lines of OCaml, while the verified part is 34,000 lines of Coq excluding comments (as counted by \codePlain{coqwc} there are 4847 lines spec +  6320 proof for AST to MiniC, and 15714 spec + 6993 proof for MiniC to EVM).
Figure~\ref{fig:compiler-overview} gives an overview on the 12~phases
of the compiler.

\subsection{Typechecking and elaboration into core language}
The DeepSEA compiler frontend works in two steps. In the first step
the dsc tool typechecks the program. In the second step it writes out
a term in the \emph{core language}---\textit{desugared} into a
functional specification in Coq. The core language is defined as a
datatype in Coq, and the desugaring function serves as a denotational
semantics which interprets the core terms as simply typed lambda
terms.

The core language syntax consists of typed expressions and commands.
Expressions are typed with respect to variable environments and commands are typed with respect to variable environments as well as layer signatures. 
Note that the distinction between expressions and commands makes it redundant to define the formal semantics for expressions. Expressions are pure
and do not have side-effects, and are just a subset of the lambda calculus. Hence the semantics for desugaring them just involve an embedding.
Since DeepSEA only supports for loops, it is possible to translate them into a functional specification using a total function (forM).
The formal semantics for the commands are defined denotationally with the help of monadic combinators \codePlain{ret, bind, get, set}:
\begin{center}
\begin{tabular}{r c l} 
	\textit{synth\_spec} (val($e$)) &=& \codePlain{ret} $e$   (an embedding of an expression) \\
	\textit{synth\_spec} (v) &=& gets ($\lambda$  $t.t$)   (gets ($f: T \rightarrow U$) := bind get (fun $x$ $\Rightarrow$ ret ($f x$)))\\ 
	\textit{synth\_spec} (v $\leftarrow$ e) &=& modify ($\lambda$ t.set \{t with $v=e$ \}) \\ 
	(modify ($f: T \rightarrow T$) &=& bind get (fun $x$ $\Rightarrow$ put ($f x$))) \\
	\textit{synth\_spec} (let $x=c_1$ in $c_2$) &=& bind \textit{synth\_spec}($c_1$) ($\lambda$ x. \textit{synth\_spec}($c_2$))\\
	\textit{synth\_spec} (if $e$ then $c_1$ else $c_2$) &=& if $e$ then\textit{synth\_spec}($c_1$) else \textit{synth\_spec}($c_2$)\\
	\textit{synth\_spec} (for $x=e_1$ to $e_2$ do $c$) &=& forM $e_1$ $e_2$ \textit{synth\_spec}($c$)\\
	\textit{synth\_spec} (assert($c$)) &=& guard \textit{synth\_spec}($c$)
\end{tabular}
\end{center}

When we load a DeepSEA-generated file into Coq, we are evaluating the
desugaring function to generate the functional specification for the
contract methods. The user proofs are done with respect to that
specification, and dsc generates a proof that the specification is
refined by the bytecode, so although the dsc tool is not itself
verified it is not in the trusted computing base.

\subsection{Translation into MiniC intermediate language}
\label{sec:minic}

Rather than translating the core AST directly into EVM bytecode, we
first translate it to an intermediate language which we call MiniC.
The syntax is shown in Figure~\ref{fig:minic-syntax}.
As the name suggests it is similar to a subset of C, with the main
differences being on the one hand a more limited support for pointers,
and on the other a set of primitive command for accessing blockchain
features such as transmitting money or querying the blockchain state.

The DeepSEA source language was designed to be quite constrained in
order to make it easy to desugar it into Coq functions. For example it
does not have mutable local variables, and loops are restricted to a
few terminating patterns. By contrast, MiniC is general-purpose, and
we believe our backend to be reusable for creating verified compilers
for other blockchain languages.

\subsubsection{Memory Model}
\label{sec:memory-model}

In MiniC, most addresses (that is, L-values) are represented as {\em
  extended identifiers}---recursively defined paths from a root
identifier through its data structure (Figure~\ref{fig:extids}).
For instance, if the identifier \(i\) refers to an array of structs containing a field identified by \(i'\), then \(\codePlain{Index}(i,0)\) points to the first struct in the array, and
\(\codePlain{Field}(\codePlain{Index}(i,0),i')\) refers to the field \(i'\) in the first array.
The {\em root} of an extended identifier is either \(\codePlain{Global}(i)\), indicating a persistent global variable, or \(\codePlain{Local}(i)\), which will be
allocated and deallocated at function calls and returns. The set of L-values also contains raw hashes. A memory \(m\) is a map from L-values to values.

Using extended identifiers in place of a more typical block-offset model gives us some advantages. First, the structure of an extended identifier
corresponds directly to that of the sequences of hashes that we use to access EVM storage at lower-level IRs, making compilation and proofs simpler.
Second, there is a clear distinction between addresses that would be aliased in the block-offset model: a pointer to an array is different from the
pointer to its first value. A pointer to an object's substructure, when passed as an argument, gives access only to that substructure, not the
object as a whole as is typical in C-like languages.

\begin{figure}[t!]
	\centering
	\begin{subfigure}[t]{0.3\textwidth}
		\centering
		\begin{grammar}
			<type> ::=  \codePlain{Tvoid}   (the [void] type)
			\alt \codePlain{Tint}($i,s$)    (integer types)
                        \alt \codePlain{Tpointer}($p, t$)    (pointer types)
                        \alt \codePlain{Tarray}($t, z$)   (array types)
                        \alt \codePlain{Thashmap}($t_1,t_2$)    ( key type, elem type )
                        \alt \codePlain{Tfunction}($ty\_list$ , $t$)     (function types )
                        \alt \codePlain{Tstruct}($i, f\_list$)    (struct types)
                        \alt \codePlain{Tunion}($i, f$)     (union types)
                        \alt \codePlain{Tcomp\_ptr}($i$)    (pointer to named struct or union)

			<ty\_list> ::= \codePlain{Tnil} 
			\alt \codePlain{Tcons} ($t, t_l$)

			<f\_list> ::= \codePlain{Fnil}
			\alt \codePlain{Fcons}($i,t,f_l$)
	        \end{grammar}
		\caption{Type representation}
	\end{subfigure}
	\hfill
	\begin{subfigure}[t]{0.3\textwidth}
		\centering
		\begin{grammar}
			<expr> ::= \codePlain{Eint} ($i,t$)  (integer literal)
			\alt \codePlain{Eint256} ($i_{256}, t$)  (256-bit integer literal)
			\alt \codePlain{Evar} ($i,t$)  (variable)
			\alt \codePlain{Eglob} ($i,t$)  (global variable)
                        \alt \codePlain{Etemp} ($i,t$)  (temporary variable)
			\alt \codePlain{Ederef} ($e,t$) (pointer dereference)
			\alt \codePlain{Eaddr} ($e,t$)  (address of an lvalue converted to an rvalue pointer)
			\alt \codePlain{Eunop} ($op, e, t$)  (unary operation)
			\alt \codePlain{Ebinop} ($op, e_1 , e_2, t_1, t_2$)  (binary operation)
                        \alt \codePlain{Efield} ($e, i, t$)  (struct member access)
			\alt \codePlain{Eindex} ($e_1, e_2, t$) 
			\alt \codePlain{Ecall0} ($b_0$)
			\alt \codePlain{Ecall1} ($b_1$)
		\end{grammar}
		\caption{Expressions}
	\end{subfigure}
	\hfill
	\begin{subfigure}[t]{0.3\textwidth}
		\centering
		\begin{grammar}
			<statements> ::= \codePlain{skip}
			\alt \codePlain{assign}($e_1,e_2$)
			\alt \codePlain{set}($i,e$)
			\alt \codePlain{call}($o,l,$ list $e$)
			\alt \codePlain{sequence} ($s_1, s_2$)
			\alt \codePlain{ifte} ($e, s_1, s_2$)
			\alt \codePlain{loop} ($s$)
			\alt \codePlain{break} 
			\alt \codePlain{return} ($o$)
			\alt \codePlain{transfer} ($e_1,e_2$)
			\alt \codePlain{callmethod} ($e,$ list $i, i_1$)
			\alt \codePlain{log} (list $e_1 ,$ list $e_2$)
			\alt \codePlain{revert}
		\end{grammar}
		\caption{Statements}
	\end{subfigure}
	\caption{MiniC syntax}
        \label{fig:minic-syntax}
	\label{MiniC}
\end{figure}
\begin{figure}
	\begin{subfigure}[t]{0.45\textwidth}
		\begin{grammar}
			<\(l \in\) L-values> ::= \codePlain{Eid} (\(p\))
			\alt \codePlain{Lhash1} ($i_{256}$)
			\alt \codePlain{Lhash2} ($l, i_{256}$)
		\end{grammar}
		\caption{Left Values}
	\end{subfigure}
	\hfill
	\begin{subfigure}[t]{0.45\textwidth}
		\begin{grammar}
                       <\(p \in\) extended identifiers> ::= \codePlain{Global}
			\alt \codePlain{Local} ($i_{256}$)
			\alt \codePlain{Field} ($p , i$)
			\alt \codePlain{Index} ($p , i_{256}$)
		\end{grammar}
		\caption{Extended identifiers}
	\end{subfigure}
	\caption{Extended identifiers as addresses}
        \label{fig:extids}
\end{figure}

\subsubsection{Datatype representation relation}
\label{sec:datatype-representation}

One of the  design decisions that makes reasoning about DeepSEA generated specifications user-friendly
is the choice of data representation in these functional counterparts of the source code. We use mathematical types available in Coq such as $Z$, finite maps etc. to represent datatypes in the theorems that reason about specifications. While that makes reasoning easier, the compiler must maintain a correspondence between these types and the conventional C types used in the MiniC AST and
ensure that this relation is preserved by any data manipulation.

The state representation for the execution context model used to reason about the specifications, is given by a record in Coq (recall the token contract example in section \ref{sec:deepsea-pl}) with a field for each object
in the DeepSEA source code containing a mathematical value. The state representation in the MiniC IR consists of the memory model, described below, which is further parametrized
by a value  $d$ of an \textit{abstract data type}, representing trusted primitives. Defining a relation between these two state representations involves defining a correspondence between the types of values
used in both the contexts and then a relation between the memory locations and the MiniC representation of these values. We describe both these relations here.

A relation of the form $p \xmapsto{m, ty} v$ holds when upon loading from the memory $m$ at position $p$ yields the value $v$ which has type $ty$.
This defines the base case for simple values, for complex values like structs and arrays the relation is defined recursively point-wise.

With this relation defined, we now define the relation between type representations and then the corresponding value representations in the specification and the MiniC implementation. The mathematical types
used in specifications are mapped to a corresponding C type via the inductively definition \codePlain{type\_pair}. This relation can be thought of as a map from mathematical types to C types, say $tp(.)$
Finally, the relation between the specification values and the MiniC values is parametrized
by the mathematical type representing the value as illustrated by a few base cases in the figure below.
\begin{figure}[t!]
\centering
	\begin{subfigure}[t]{0.3\textwidth}
		\centering
		\textit{tp}(\codePlain{int}) = \codePlain{Tint} \\
		\textit{tp}(\codePlain{unit}) = \codePlain{Tvoid}
	\end{subfigure}
	\hfill
	\begin{subfigure}[t]{0.3\textwidth}
		\centering
		\textit{R}$_{int} \text{ }n \text{ }$\codePlain{Vint}$\text{ n } \iff$\\
		$0 \le n \le 2^{32}$\\
		\textit{R}$_{unit} \text{ ()} \text{ }$\codePlain{Vint}$\text{ n }$ $\iff$\\
		$(n = 0)$
	\end{subfigure}
	\hfill
	\begin{subfigure}[t]{0.3\textwidth}
		\centering
		\begin{prooftree}
			\AxiomC{\textit{R}$_{T_i}\text{ }t.f_i \text{ }v_i$}
			\AxiomC{$p_i \xmapsto{m, tp(T_i)}v_i$ }
			\BinaryInfC{$t\mathcal{R}(m,a)$}
		\end{prooftree}
	\end{subfigure}
	\caption{Data representation and State relations}
\end{figure}

Once we have these two relations in place, they are composed for each object field $f_i$ in the record representing a layer state for the specification and a corresponding extended identifier $p_i$ in 
the MiniC program, to define the relation $\mathcal{R}$ for the entire layer.

\subsubsection{Compilation correctness theorem}
\label{sec:correctness-statement}
The command synthesis rules for this phase are fairly standard, and we omit them here while making one remark.
Note that because object fields behave differently when loading values and assigning them respectively, expressions are further classified as L-values and R-values respectively, and have distinct
translation rules for both.

Finally, there is a correctness theorem proved in Coq which states that the command translation is correct with respect to desugaring \textit{if} the verification
conditions regarding data representation generated by dsc hold. For
each command $c$ the DeepSEA system defines a condition
$\mathcal{VC}(c,t)$ that states $c$ can be correctly executed in state
$t$. This is a large conjunction generated based on the datatypes
involved which the user must prove (with the aid of some provided
tactics); for example each time an integer variable is assigned the
user must prove that it is in the range $0 \le x < 2^{256}$.  For a detailed exposition on
the definition of such verification conditions, and the tactic
mechanism designed to solve them, we refer the reader to a previous
publication~\cite{sjoberg:deepsea}.

With the verification condition aspect of this compiler duly noted, we have the compiler correctness theorem
stated as a simulation:
\begin{theorem} For every DeepSEA command $c$, its MiniC implementation $m(c)$, its specification $s(c)$ , a state $t$ such that $\mathcal{VC}(c,t)$ holds and a state $t'$ reached from $t$
	upon the execution of $\mathsf{runStateT}\ s(c)\ t =
        \mathsf{Some}(t')$, if we have $t \mathcal{R} (m,a)$ for some
        memory state $m$, then there exists some amount of gas $g$ and
        a memory $m'$ reachable from $m$ by execution of $m(c)$ such that 
	$t' \mathcal{R}(m',a)$ holds.
\end{theorem}

It's worth noting that this correctness theorem neatly separates out
three different source of undefined behavior. First, runtime errors
or assertion failures cause a transaction to revert; this is
represented by the high-level specification returning $\mathsf{None}$
instead of the $\mathsf{Some}$. An attacker can always set up
situations where a contract reverts, so the compiler will ensure that the
compiled bytecode faithfully follows the source program semantics in
this case.

Second, it is the programmer's responsibility to avoid
integer overflows and similar sources of overflow: DeepSEA will the
make the user prove that there are none in the $\mathcal{VC(c,t)}$
condition, and then the compiler can assume that no such behavior is
present. In this way, a DeepSEA program can be compiled more
efficiently than a Solidity program using the ``safe math'' library,
because the latter will insert a runtime check after \emph{every}
arithmetic operation, while the DeepSEA programmer only needs to
insert as many checks as needed for the $\mathcal{VC}$-proofs to go
through.

Third, out-of-gas errors are treated separately. When proving safety
properties gas cost can be ignored, and indeed it is not part of the
current generated specifications $s(c)$. In future work, we may extend
the system to also generate a second set of specifications which
exactly tracks the gas usage of each command. 

\subsection{Verified compiler backend}

The backend for the MiniC language has two compilation paths, compiling to either EVM or ``Ethereum-flavored Web Assembly'' (eWasm), but the eWasm path does not yet have a correctness proof. The EVM path consists of a series of intermediate representations, ending with a final phase that connects this IRs to a formalization of the EVM written in Lem and extracted into Coq. This formalization~\cite{lem_ethereum} of the EVM has been tested against the VM test suites provided by the Ethereum Foundation.

The source language MiniC and the first few phases of the compiler are inspired by Comp\-Cert~\cite{leroy2016compcert}, but the later phases are quite different because we target a stack machine.
Similar to CompCert, the correctness of compilation theorem for each pass is a forward simulation diagram, which is in the form of a lock-step, plus or star simulation. The proofs can then be composed into a simulation for the entire compiler.
Because the source language is deterministic, the forward simulation proof also implies a bisimulation relation.

\subsubsection{General Framework}
In order to define the semantics of all the intermediate languages, all phases of the backend share certain data. First, a \emph{global environment} which maps function names to function definitions and global variable names to the corresponding locations. Second, to support EVM operations such as querying current account balances there is a \emph{machine environment}, a record which contains information about the contract
and state of the VM. Finally, following CompCertX~\cite{gu:dscal}, the backend and the machine environment are further parametrized by an abstract data type called \textit{adata}. This is a way to support separate compilation: in addition to the concrete memory each program state also contains an abstract data value, and when compiling a particular function this value is instantiated to a record which describes the state of other functions and contracts. A global environment along with an initial memory and abstract state are constructed at the beginning of the program execution.

The syntax for the first seven intermediate languages consists of statements and expressions. The distinction being that expressions do not have side-effects while statements might.
\textit{Expressionless} is the last IR which deals with expressions.

Each compilation pass manipulates values which are represented throughout the backend by the following type:
\begin{minted}[fontsize=\footnotesize, linenos, breaklines]{coq}
Inductive val: Type := (* defined in LowValues.v *)
| Vunit: val
| Vint: int256 -> val
| Vhash: val -> val
| Vhash2: val -> val -> val.
\end{minted}
Programs for the EVM make very frequent use of 256-bit hash operations, and generally assume that there are never any hash collisions; for example all hash mappings are stored directly with no attempt to detect collisions. Therefore we use this symbolic representation of constructing a hash value from one key
and constructing a hash value from two keys respectively, which implicitly ensures that all hash operations are injective. In the final compiler phase we need an axiom (which is actually false, but ``true enough'') stating that the concrete Keccack hash function is injective.

The operational semantics for each intermediate representation are defined as labeled transition systems. The
transition relation takes the form $G \vdash S_1 \xrightarrow[]{\text{s}} S_2 $. Here $G$ represents the general environment
on which each compilation pass is parametrized (the global environment \codePlain{genv} and the machine environment \codePlain{machine\_env}), $S_1 , S_2$ 
represent the states of the transition system and $s$ represents one step of execution from state $S_1$ to $S_2$. 
The semantics of each such step corresponding to an instruction in the code is defined for each compilation pass by the inductively defined relation \codePlain{step}.
What constitutes the program state varies for each intermediate language, but for most of the passes they come in four different kinds:
\begin{itemize}
	\item regular states (\codePlain{State}) represent the states of the program during
		execution of steps which do not require control to transfer to another function;
	\item call states (\codePlain{Callstate}) represent the state of the program when it performs
		a call to another function within the contract;
	\item initial states (\codePlain{Initialstate}) represent the state of the transition system
		before contract execution begins;
	\item return states (\codePlain{Returnstate}) represent the state of the transition system when a function is about to return.
\end{itemize}

\paragraph{Handling gas}
\label{sec:gas-handling}
Because  we are compiling to EVM, each state type also contains a natural number which is the amount of gas the program has consumed, and which is incremented as each instruction is executed. The simulation relation for each phase specifies that the target code consumes at most as much gas as the sources. The amount consumed may go down, either because the compiler optimizes the code, or because we deliberately overapproximated e.g. the gas cost of a function call (by assuming the maximal number of arguments) to make the source language semantics simpler for the end user.

This convention for gas handling is the opposite of the actual EVM, and the EVM formalization we target: there the state counts the remaining gas, which is decremented by each instruction, and the program becomes stuck/reverts if the count hits zero. In early versions of the backend we tried the same, but the counting-down convention does not let allow a strong enough correctness theorem. We want to say that, if the source program computes a value using some amount $g$ gas, then running the target program will either compute the same value or run out of gas (and if you submit at least $g' > g$ gas it will successfully compute). An attacker can force a program to run out by paying for too little gas, so we want to know that the compiled contract will still end up in a safe state in that case. But because of optimization, the compiled program may not run out of gas even if the source semantics claims it will, so we do not want a source semantic that claims that programs will always abort if the gas reaches zero; we want the source program to continue executing to know what it \emph{would} have computed. At the final EVM generation compiler phase there are no further optimizations so we can  prove an exact relation between the ``gas used'' and ``gas remaining'' counts. 

Next we give a brief description of some  of the important phases of the Ethereum backend of the compiler. 

\subsubsection{Clike}
The Clike phase models persistent storage separately from memory, instead of keeping it as part of the extended environment. In particular, when compiling accesses \codePlain{a[k]} to a hashmapping in storage, this phase generates the actual hash operation.

The Clike IR shares the same syntax as MiniC, but the semantic evaluation rules for expressions differ.
Left-expressions are no longer implicitly 
dereferenced, and instead evaluate to pointers which must be explicitly dereferenced. Furthermore, the storage environment is not the same as the extended environment. It corresponds to the EVM's storage 
and hence the keys are represented by an inductive type $HashKey$ which represents an abstraction of hash values
as opposed to extended identifiers:
\begin{minted}[fontsize=\footnotesize, linenos, breaklines]{coq}
Inductive HashKey :=
| singleton : int256 -> HashKey
| pair : HashKey -> int256 -> HashKey.
\end{minted}

The phase leaves statements the same but changes the expressions embedded in them.
Since the storage is modeled explicitly in this pass, the $Eglob$ expressions are also translated to first evaluate the identifiers as hash values.
Other expressions dealing with extended identifiers, such as $Eindex$ and $Efield$, are treated based on the type of the left expression: if it is a memory pointer it is unmodified while if it is a storage pointer it is evaluated as a hash value using the $Ebinop$ expression type.

In order to define a simulation relation between the states of the two IRs, we first define relations to match the storage and the extended environment, ensure that no additional gas is consumed in this translation and match every other component in the state representation in MiniC to its translation. 

\subsubsection{Cgraph}

The Cgraph IR represents functions as control-flow graphs, where each statement is a node with an edge to its successor.
Immediately after constructing a CFG-representation of the program, we
carry out a live variable analysis of all the temporary variables, and
use a standard graph-coloring register assignment algorithm to rename
temporary variables into a smaller set. This is important because
temporaries will eventually be mapped to items on the EVM argument
stack, and only the top 15 variables can be accessed. Other EVM compilers,
such as Solidity, give an error if a function has more than 15 local
variables, but we can accommodate more if their live ranges do not overlap.

\subsubsection{Cbasic}

The Cbasic IR groups lists of statements into blocks. Program execution happening within such a block is represented by an additional execution context constructor called \codePlain{Block}.
The list of statements enclosed within the block are syntactically the same as the statements in Cgraph. This intermediate language and phase is similar to the similarly named phase in CompCert, except that blocks can end with \codePlain{Srevert} (which aborts the entire transaction) as well as \codePlain{Sdone}.

The \codePlain{Block} constructor represents the execution context when the control is within a block of statements.
The transition step semantics for this IR are given by the inductively defined proposition \codePlain{step}, like in the previous IRs.
However, due to the additional state description given by the \codePlain{Block} constructor, the step semantics are worth mentioning.
Except for the semantics for control-flow statements where the execution context transitions from a \codePlain{Block} to a \codePlain{State},
any execution of other statement only causes transition within the \codePlain{Block} state, where the change is that a statement is consumed.
The execution of the \codePlain{Scall} statement on the other hand, causes the state to transition from a \codePlain{Block} to a \codePlain{Callstate}.

The translation of statements from the Cgraph IR to a block in the Cbasic IR is done by appending a \codePlain{Sjump node} statement at the end of each statement from Cgraph except \codePlain{Srevert} and \codePlain{Sdone}, to obtain a bblock. 

\subsubsection{Clinear}
This pass linearizes the CFG generated from the Cbasic IR. The code in a program execution is now a list of statements instead of a DAG.  The CFG structure aids 
controlflow analysis, but we eventually need to print out a linear list of assembly instructions. The syntax for the expressions remains the same, and
the statements in this IR largely remain the same, except for two new additions \codePlain{Sfetchargs, Sintro}. These statements push data onto the call stack corresponding
to the arguments for the EVM calldata operation and adding fresh temporaries respectively.

The code linearization process begins with an enumeration of all the labels that appear in the body of the function we are compiling. There is a heuristic that generates them using a depth-first traversal of the CFG, in order to avoid inserting unnecessary jump instructions between basic blocks.
Once we have a list of nodes, the code for this phase is generated from the code in Cbasic by concatenating the blocks in Cbasic in the order of the enumeration of the
labels. 

The \codePlain{match\_states} relation is defined using a \codePlain{match\_stackframes}
relation between the calling context in the source and target languages, and mapping function representations to their translated counterparts. Certain subtleties of this relation come through while defining it for the execution contexts for the custom
Ethereum statements. We state the case for the \codePlain{Stransfer} statement failing here. The cases for conditional jumps (\codePlain{Sjumpi}) and method calls (\codePlain{Scallmethod}) failing
are treated similarly and so we omit them for brevity.
\begin{itemize}
	\item \lstinline{match_states_transfer_fail} : This constructor is defined to relate execution of the \textit{Stransfer} statement which is an EVM specific statement. It models the case when 
		the execution fails. Similar to the constructor to model successful conditional jumps, the state reached in the Cgraph execution context once the transfer fails is matched to the
		state in the Clinear execution context where the condition for the transfer to fail holds, but the step is yet to be executed. The condition expressing this failure is stated as:
		\begin{minted}[fontsize=\footnotesize, breaklines]{coq}
DOJUMP: (MachineModelLow.me_transfer me) a' v' lg' lg Int256.zero)
		\end{minted}
		Hence, the state
                \begin{minted}[fontsize=\footnotesize, breaklines]{coq}
(State s f pc (TempsL le) se lg g)
                 \end{minted}
                 is matched to the state
                 \begin{minted}[fontsize=\footnotesize, breaklines]{coq}
(State ts tf (Stransfer a v pc :: c) (TempsL le) se lg' (g' + gas_transfer_cintptr a v 0)).
                 \end{minted}
\end{itemize}
 The simulation diagram for the correctness proof for this translation is a star simulation, and hence we need a notion of measure to prevent the problem of stuttering. After much
 experimentation, the measure on the Cbasic state representations is defined over non-negative integers with all states except the \codePlain{Block} (which has measure 1) state having measure 0.
\subsubsection{Stacked}

This is the first IR where the syntax for the expressions is different from the MiniC expressions. The syntax for this IR is given in Figure~\ref{stacked-statements}.
The code is represented as a list of statements and functions are again record types, but this time with only one field \codePlain{fn\_code}.
The function \textit{stacked\_expr} converts expressions from Clabeled to a list of statements in Stacked. The semantics of expressions from the previous phase on whether they are to be treated as
R-values or L-values is conveyed through a boolean which is passed as an argument to \textit{stacked\_expr}. This value is true if the expression is to be evaluated in a right handed context and false otherwise.
The state of the execution context is represented in a similar way as the previous pass. One notable change is that instead of 
having a callstack and a local environment, every state now has \codePlain{stack\_entries}. These \codePlain{stack\_entries} are represented as a list of the tagged type \codePlain{LowValue.val + generic\_label}. Values pushed as \codePlain{stack\_entries} represent the values that expressions are evaluated to and the \codePlain{generic\_label} data type represents typed labels which are annotated with the function from
which they were pushed. This is done so that the function to be returned to can be looked up from the general environment. 
The value of the expression that it evaluates to and how it modifies \codePlain{stack\_entries} is dictated by the \codePlain{eval\_rvalue} function, whose excerpt we give here:
\Needspace{100pt}
\begin{minted}[fontsize=\footnotesize, linenos, breaklines]{coq}
Inductive eval_rvalue: expr -> stack_entries -> stack_entries -> Prop :=
    | eval_Econst_int256: forall i st,
        eval_rvalue (Econst_int256 i) st (inl (LowValues.Vint i) :: st)
    | eval_Etempvar: forall i v st,
        stack_get st i = Some v ->
        eval_rvalue (Etempvar i) st (inl v :: st)
    | eval_Esload: forall ptr i st,
        HashEnv.read ptr he = Some i ->
        (*read' ptr he = Some (OfLow i) ->*)
        eval_rvalue Esload (inl ptr :: st) (inl i :: st).
\end{minted}
\begin{figure}[H]
	\centering
	\begin{subfigure}[t]{0.4\textwidth}
		\centering
	        \begin{grammar}
		<expression> :: = \codePlain{const256}($i$)
		\alt \codePlain{global}($i$)
		\alt \codePlain{temp}($n$)
		\alt \codePlain{sload}
		\alt \codePlain{unop}($op$)
		\alt \codePlain{binop}($op,b$)
		\alt \codePlain{call0}($b0$)
		\alt \codePlain{call1}($b1$)
		\end{grammar}
		\caption{Expressions}
	\end{subfigure}
	\hfill
	\begin{subfigure}[t]{0.4\textwidth}
		\centering
		\begin{grammar}
		<statement> :: = \codePlain{skip}
		\alt \codePlain{rvalue}($e$)
		\alt \codePlain{pushvoid}
		\alt \codePlain{pop}
		\alt \codePlain{assign}
		\alt \codePlain{set}($n$)
		\alt \codePlain{done}($n, r_t$)
		\alt \codePlain{pushlabel}($t$)
		\alt \codePlain{label}($l$)
		\alt \codePlain{jump\_call}
		\alt \codePlain{jump\_internal}
		\alt \codePlain{jumpi}
		\alt \codePlain{transfer}
		\alt \codePlain{callmethod}($i, n_1, n_2$)
		\alt \codePlain{log}($n_1,n_2$)
		\alt \codePlain{revert}
		\alt \codePlain{fetchargs}
	\end{grammar}
		\caption{Statements}
    \label{stacked-statements}
	\end{subfigure}
	\caption{Stacked syntax}
	\label{stacked}
\end{figure}
The state of the execution context is parametrized by the arguments to be passed to the \codePlain{Scallmethod} statement. 
They are no longer a part of the arguments that the statement takes. Instead, the statement
is passed the number of arguments that it needs. Since these arguments do not change throughout the program execution, they are declared as a parameter instead of being passed around in the state. 
The translation from Clableled to Stacked is parametrized by a map of temporary variables mapped to their values. This map is initialized by the temporaries declared as a part of the function definition
in Clabeled using the \codePlain{allocate\_fn\_temps} function. Once this map is initialized, it is passed as the parameter for translating Clabeled code to Stacked code, which is then used to construct
a Stacked function from a Clabeled function.
Finally, the \coq{stacked_genv} function converts the global environment from Clabeled to the global environment for Stacked using the functions listed above to translate the corresponding parts.
The whole program is converted from its representation in Clabeled to that in Stacked after performing a check that the function and method labels do not repeat.
The semantic preservation of the translation is proved using a plus simulation diagram.
\subsubsection{Expressionless}
As is evident from the name, this is the first IR where expressions are no longer a part of its syntax. The syntax for statements in this IR is given in Figure~\ref{fig:expressionless-syntax}.

\begin{figure}
	\centering
	\begin{grammar}
        <statement> ::= \codePlain{push}($v + l$)
		| \codePlain{dup}($n$)
		| \codePlain{sload}
		| \codePlain{unop}($op$)
		| \codePlain{binop}($op,b$)
		\alt \codePlain{call0}($b_0$)
		| \codePlain{call1}($b_1$)
		| \codePlain{skip}
		| \codePlain{pop}
		| \codePlain{sstore}
		| \codePlain{swap}($n$)
		| \codePlain{done}($r_t$)
		\alt \codePlain{label}($l$)
		| \codePlain{jump}
		| \codePlain{jumpi}
		| \codePlain{transfer}
		| \codePlain{callmethod}($i, n_1, n_2$)
		\alt \codePlain{log}($n_1,n_2$)
		| \codePlain{revert}
		| \codePlain{calldataload}
		| \codePlain{constructordataload}($n$)
	\end{grammar}
  \caption{Expressionless syntax}
  \label{fig:expressionless-syntax}
\end{figure}

The code, function, global environment and program representation are the same as in the Stacked IR. It is from this pass where the transition from the CompCert style
backend to an IR written to match the execution context of the EVM becomes clear. Since the EVM has a special function called the constructor, which compiles differently
than other methods in the contract, it is compiled in a different manner than normal methods/functions. In order to indicate whether
a particular function is a constructor a datatype \codePlain{function\_kind} is defined and is passed around as the type of the function.
The compilation of a constructor differs from that of a normal function in two primary ways. 
Firstly, function calls from the constructor are prohibited and second is that the 
callmethod arguments which parametrize the execution context of this IR (similar to Stacked) are loaded in a different way for constructors. The state representation for this IR 
is the same as that for Stacked except that there is no separate \codePlain{Callstate} for function calls.
The function \codePlain{expressionless\_stm} takes a stacked statement, a \codePlain{function\_kind} argument to indicate the type of function for which the compilation is going on (needed to determine what \codePlain{Sjump\_call} gets compiled to) and returns an 
option type on list of expressionless statements. Every translation function takes an argument
of the type \codePlain{function\_kind} to determine what type of function is being compiled and returns a translation accordingly.
Finally, the \codePlain{expressionless\_genv} function is used to construct the global environment, and to convert the program representation
from Stacked after performing the check that labels in the program do not repeat and each function begins with it's own label.
\subsubsection{Methodical}
The syntax for this IR is the same as that for Expressionless. The only and major difference is in the semantics where this IR instead of 
providing separate entry points into the contract for each method,
streamlines the contract code into one list of statements with a single entry point into the contract that ``multiplexes'' into the correct method. This is reflected as expected in the function that translates the Expressionless program to return the code for execution 
in the Methodical IR. This is done by the \codePlain{methodical\_main} function which uses the helper function 
\codePlain{methodical\_multiplexer\_body}. \codePlain{methodical\_multiplexer\_body} is a recursively defined function that takes the list of methods and a map which takes a method to its definition as arguments,
and returns an option type on code. For each method in the list, it looks for its definition in the map and returns a list of statements which are semantically equivalent to performing a conditional jump to the method definition if the signature of the function
that is to be executed matches the method label. This is the main step for this compilation pass which achieves the effect described in the
paragraph above.

\subsubsection{EVM\_lem}

This is the final pass of the DeepSEA compiler backend. The methodical IR is translated to the syntax for the execution of code in this EVM formalization as defined in \cite{lem_ethereum}. This formalization acts as an interpreter for a single smart contract execution. This is achieved by providing a formal definition of the EVM which is widely portable. Since all smart contract executions on the EVM happen within a 
block of various transactions clumped together by the miner, the execution context here is a block which is defined using a Coq record data type \codePlain{block\_account}.
The execution context for this phase is completely characterized by two record types, the
\codePlain{constant\_ctx} and \codePlain{variable\_ctx}. The \codePlain{constant\_ctx} consists of fields which represent relatively stable data, unaffected by the execution of instructions but needed
to access the information required to successfully execute bytecode. The \codePlain{variable\_ctx} on the other hand constitutes of fields which represent data susceptible to change with instruction
execution. 
The program is also represented using a record type:
\begin{minted}[fontsize=\footnotesize, linenos, breaklines]{coq}
Record program : Type := {
  program_content : Z  ->  option  inst  ; (* a way to look up instructions from positions *)
  program_length  : Z  (* the length of the program in bytes *)
}.
\end{minted}
The equivalent of the statement datatype from the previous passes is the \codePlain{inst} datatype which is an inductively defined wrapper type used to represent EVM opcodes. The possible outcomes
of executing an opcode are captured by the \codePlain{instruction\_result} datatype defined as follows:
\begin{minted}[fontsize=\footnotesize, linenos, breaklines]{coq}
Inductive instruction_result : Type :=
| InstructionContinue:  variable_ctx  -> instruction_result  (* the execution should continue. *)
| InstructionToEnvironment:
  (* the execution has stopped; either for the moment just calling out another account, or *)
  (* finally finishing the current invocation *)
    contract_action    (* the contract's move *)
  ->  variable_ctx       (* the last venv *)
  ->  option  ((Z * Z) % type)  -> instruction_result .
\end{minted}
A total function \codePlain{instruction\_sem} defines the semantics of all the opcodes and their effect on the state of the EVM. A helper function \codePlain{meter\_gas} computes the amount
of gas required for the execution of the opcode and then \codePlain{subtract\_gas} subtracts it from the gas balance of the contract. 
Finally, the execution of the entire program and its effects on the state of the EVM is defined using a recursive function \codePlain{program\_sem}:
\begin{minted}[fontsize=\footnotesize, linenos, breaklines]{coq}
Program Fixpoint program_sem  (stopper : instruction_result -> unit) (c : constant_ctx) (fuel : nat) (net : network) (pr : instruction_result)  : instruction_result :=
match ( fuel) with
 | 0%nat => pr
 | S (fuel_pred) => program_sem stopper c fuel_pred net (next_state stopper c net pr)
end.
\end{minted}

Here the function \codePlain{next\_state} is a helper function which acts on the EVM state in the same way as the fold function does on lists. The program execution for this pass is modeled in the form of a game between the environment and the contract. The possible moves of both the players are defined via the inductive types \codePlain{environment\_action}
and \codePlain{contract\_action} respectively. The operational step semantics are defined by the \codePlain{step} definition which takes a \codePlain{global\_state} as an argument and returns the modified \codePlain{global\_state} obtained upon execution of an operation. 
The record type \codePlain{global} completely characterizes the VM state including a list of ongoing calls and their respective 
execution states.
For a detailed exposition of the model used in this formalization, we refer the reader to~\cite{lem_ethereum}.
The translation of statements from the Methodical IR to the syntax in the EVM formalization is done via a total function 
\codePlain{s\_t\_opcodes} which takes as arguments a \codePlain{statement},
and a map from labels to integers (this is needed in order to convert labels being pushed onto the stack into bytes) and 
returns a list of instructions.

In order to prove semantic preservation for this pass, we set up the correspondence between different components of the execution 
context in the following way:
the \codePlain{block\_account} corresponding to the execution context in the Methodical phase is constructed through the \codePlain{initial\_state\_account} definition which takes the account
address and the storage (which is defined to be a function that takes a word of size 256 and returns the corresponding word in the storage of that size) as arguments and returns the corresponding \codePlain{block\_account}.
Similarly, the initial variable and constant context (\codePlain{variable\_ctx}, \codePlain{constant\_ctx} respectively) are constructed as functions of the account address, storage and the amount of gas available for program execution. The definitions \codePlain{constant\_ctx\_init} 
and \codePlain{variable\_ctx\_init} are defined for this purpose.
The definition \codePlain{initial\_arguments} initializes the environment with the appropriate calldata arguments and the constructor code used to deploy the runtime code for the contract.
Using these definitions, \codePlain{initial\_global} takes the account address, the storage and the remaining gas as arguments and constructs an instance of the \codePlain{global} record type described above.
Once the environment has been initialized with the appropriate data, we further setup the variables to correspond to the entry point to the code of the methodical pass. This is done via the \codePlain{enter\_code\_account}, \codePlain{enter\_code\_cctx}, \codePlain{enter\_code\_vctx},
\codePlain{enter\_code\_state} and the \codePlain{enter\_code\_global} definitions respectively. Similar helper definitions are defined to
relate the final states in both the languages. 
In order to set up the simulation relation to prove the semantic preservation, we define a wrapper proposition to mimic the step semantics in the methodical pass:
\begin{minted}[fontsize=\footnotesize, linenos, breaklines]{coq}
Inductive lem_step (n : network) : option global_state -> option global_state -> Prop :=
 | Init_state : forall addr he l g,
   lem_step n None (Some (Continue (enter_code_global addr (hash_env_to_store he) g l)))
 | Take_step : forall g1 g2,
   block.step n g1 = g2 ->
   lem_step n (Some g1) (Some g2).
\end{minted}
Since the code is represented as a map as opposed to a list of statements, in order to define a relation between the respective state representations, we need to relate the corresponding elements of the states. The characterizing elements are the code in the account, the storage, the memory and the stack. Here we present the correspondence for each of these in order:
\begin{itemize}
	\item \emph{Code representation}:
               we first define an inductive relation to match the program counter in the Lem formalization to the corresponding statement 
		being executed in the methodical IR:
\begin{minted}[fontsize=\footnotesize, linenos, breaklines]{coq}
Inductive code_from_counter : Z -> evm.program -> list inst -> Prop :=
| N : forall (pc : Z) (p : evm.program), code_from_counter pc p nil
| L : forall (l : list inst) (i : inst) (p : evm.program) (pc : Z) ,
      (program_content p) pc = Some i ->
      code_from_counter (pc + (inst_size i)) p l ->
      code_from_counter pc p (i :: l).
\end{minted}
                The above relation is simply a recursive way of relating valid program counter positions to the instructions that 
		can be read from them. Once we have this relation, we use it to define a relation between the respective 
		code representations as follows:
\begin{minted}[fontsize=\footnotesize, linenos, breaklines]{coq}
Definition rel_code (l_map:PMap.t Z) (p:evm.program) (code:list statement) (pc:Z) : Prop :=
  code_from_counter pc p (flatmap (fun s1 => s_t_opcodes s1 l_map) code).
\end{minted}
                The above relation simply states that the entire code in the methodical state is related to the beginning 
		position of the program counter for that code.
	\item \emph{Memory representation}:
	      this relation is defined so as to compare the values read from the memory representations in the two passes from the same
	      address. Since the datatypes used to represent corresponding values is different, we use appropriate conversion functions
		to be able to compare them.
\begin{minted}[fontsize=\footnotesize, linenos, breaklines]{coq}
Definition rel_mem (l_map : PMap.t Z) (stk :stack_env ) (m :memory ) :Prop :=
forall (i : LowValues.val) (b1 : LowValues.val), 
StackEnv.read_from_base i stk = Some b1 ->
cut_memory_aux_alt (entries_to_w256 (inl i) l_map) 32 m = val_to_list_bytes b1.
\end{minted}
     \item  \emph{Storage representation}:
	     the relation to match corresponding storage models is defined in a similar manner as the memory representation relation.
\Needspace{40pt}
\begin{minted}[fontsize=\footnotesize, linenos, breaklines]{coq}
Definition rel_store (l_map : PMap.t Z) (he :hash_env) (s:storage) : Prop := 
forall (v : val) (b1: LowValues.val), 
HashEnv.read v he = Some b1 ->
s (entries_to_w256 (inl v) l_map) = entries_to_w256 (inl b1) l_map.
\end{minted}
     \item  \emph{Stack representation}:
	    in both the languages  the stack representations are related by a recursively defined relation, which relates empty stacks
             as the base case, and relates two stacks if the top elements represent the same value and the tail stacks are related.
\begin{minted}[fontsize=\footnotesize, linenos, breaklines]{coq}
Inductive rel_stk (l_map : PMap.t Z): stack_entries -> lem_stack -> Prop :=
| Empty : rel_stk l_map ([]) ([])
| Push_elem : forall s1 s2 v1 v2 ,
        rel_stk l_map s1 s2 ->  
        entries_to_w256 v1 l_map = v2 ->
        rel_stk l_map (v1 :: s1) (v2 :: s2).
\end{minted}
      \item  \emph{Gas relation}:
             the Lem defined EVM model semantics subtract gas upon execution of each instruction, whereas the Methodical IR accumulates
	     the amount of gas required for the successful execution of the whole program. To relate the two ways of accounting for gas,
	     we require that after each instruction execution the sum of the gas components in both the state representations does not change.
	     This is because the amount that gets added in the Methodical semantics is the exact amount that gets removed in the EVM semantics,
	     thus keeping the total constant.
\end{itemize}
Now we finally define a relation for the corresponding state representations using the above relations:
\begin{minted}[fontsize=\footnotesize, linenos, breaklines]{coq}
Inductive rel (gas_limit : Z) (l_map : PMap.t Z) : state -> option global_state -> Prop :=
 | I : forall s g d,
       rel gas_limit l_map (Initialstate empty_stack s d g) None
 | E : forall s g ,
       rel_stk l_map (state_to_stack s) (vctx_stack (ext_vctx_state g)) ->
       rel_mem l_map (state_to_senv s) (vctx_memory (ext_vctx_state g)) ->
       rel_store l_map (state_to_henv s) (vctx_storage (ext_vctx_state g)) ->
       rel_code l_map (cctx_program (g_cctx g)) (state_code s) (vctx_pc (ext_vctx_state g)) ->
       vctx_gas (ext_vctx_state g) + Z.of_nat (state_gas s) = gas_limit ->
       rel gas_limit l_map s (Some (Continue g)).
\end{minted}
The correctness of the above relation is proved using a plus step simulation.

\newcommand{\ppltr}{{\sf ppltr}}
\newcommand{\ebso}{{\sf ebso}}

\subsection{Optimizations}
\label{sec:optimizations}

The Expressionless phase has machinery for performing \emph{peephole
  optimizations}. Optimization is paramount: the cost for execution is
directly derived from the executed instructions and access to
storage/memory and for every inefficiency the caller looses
money---for every call: \cite{2020_albert_et_al}
super-optimize a function within the AirdropToken smart contract,
which is called more than half a million times. In total they estimate
savings of \$2815. \cite{2020_brandstatter_et_al} analyzed optimizations of ~3000 Solidity
contracts from \texttt{etherscan.io} with respect to well-known
compiler optimization strategies and found that around 7~\% of
contracts could be improved.

In peephole optimizations, we replace small sequences of EVM bytecode
by cheaper, observationally equivalent code using rewrite rules. They
are local, closed under context, and modular, as well as adjustable by
adding or removing rewrite rules.
To obtain rewrite rules, we leverage the tool
\ppltr~\cite{2020_schett_et_al}, which generates peephole optimization
rules from optimizations found by the EVM bytecode
superoptimizer \ebso~\cite{2019_nagele_et_al}, which encodes the search
for cheaper, observationally equivalent, EVM bytecode into an SMT
problem.
While these optimizations are found by an automated theorem prover and
correct by construction with respect to the model encoded in SMT,
the correctness argument needs to be proved within the
compiler. Reproving the correctness also widens the scope: the state
modeled in SMT does not capture all of the EVM state, e.g., it does not
model memory, which can then be shown as unaffected.
Our first approach to extract SMT proofs to Coq failed due to
uncertainty on how to relate the encodings of the EVM semantics in the
two settings. The practical solution is to extend the tool \ppltr\ to
generate Coq proofs with tactics.
In our formalization, peephole optimization rules have two components:
(1) the \emph{definition} \texttt{rule\_rewrite} of the rule, and (2)
a \emph{proof} \texttt{rule\_correct} showing that if there is an
optimization \texttt{p} to \texttt{po} then after executing \texttt{p}
and \texttt{po} the global environments coincide apart the from the
gas costs.
The proofs employ tactics to execute the original
instructions~\texttt{p}, and the optimized instructions~\texttt{po}
and observe that the state is equivalent. One key lemma is to show
that gas consumption grows monotonically when executing a program. For
stack-based optimization rules, such as optimizing \texttt{DUP SWAP}
to \texttt{DUP}, equivalence of the states follows directly from the
execution.  Optimization rules targeting the algebraic properties of
instructions, e.g., zero is the neutral element in addition, need a
lemma data base.
The peephole optimization rules can be applied to any program
exhaustively. The (strict) gas decrease in every rule immediately
gives a reduction order. Then correctness of applying peephole
optimizations follows by induction from \texttt{rule\_correct} of the
individual rules. The current state of the work can be found at: 
\url{https://github.com/mariaschett/ppltr/tree/ace6324916534fc37290c870da47171fdbe4a98c}.

\section{Related work}

\paragraph{Verified compilers}
In the last decade, several large and mature verified compilers, such
as CompCert~\cite{leroy2016compcert} (which the DeepSEA compiler is partly based upon) and CakeML~\cite{tan2016new}, have
been developed.
However, the only verified compiler for smart contracts, which we are
aware of, is Elle~\cite{alvarezelle}. While Elle targets the same Lem
formalization of EVM as us, the Elle source language is less ambitious
than DeepSEA: it compiles a version of the LLL (``low level
language'') programming language. LLL is similar to a macro assembler
or an intermediate language such as Yul: it provides structured
control flow like while-loops and if-statements, but apart from that
all commands are raw EVM assembly instructions directly affecting
storage and memory; there are no higher-level data types as in
DeepSEA. Elle does not distinguish undefined behavior such as
overflows in DeepSEA, and the compiler does no optimizations.

\paragraph{SMT-based \contract{} verification}
Various tools exist for \contract{} verification that rely on SMT
solvers to discharge proof obligations. These include
Mythril~\cite{mythril_repo}, Oyente~\cite{luu2016making_oyente},
Solc-verify~\cite{hajdu2019solc}, Solidity's
SMTChecker~\cite{alt2018smt}, Verisol~\cite{born2020formal} and
Verx~\cite{permenev_verx}. A key difference between these tools and
our approach is that the DeepSEA is foundationally verified. In
addition, our use of Coq allows the full power of interactive theorem
proving which is critical for the verification of complex properties
of \contracts{} such as the Automated Market Maker. While SMT solvers
will typically solve the goals they are able to solve more quickly
than an equivalent DeepSEA proof, we believe that the trade-off for
more expressiveness is worthwhile in many cases.

\paragraph{Low-level language formalizations}
The low-level stack based \contract{} language Michelson of the Tezos blockchain is formally specified in Coq. This formalization, Mi-Cho-Coq~\cite{bernardo2019mi}, enables verification of Tezos \contracts{} at a low-level. Their work differs from DeepSEA firstly in that it does not target Ethereum, which at the time of writing is by a large margin the most popular blockchain with fully-featured \contracts{}. Secondly, DeepSEA works with a high-level functional representation of a \contract{} rather than the low-level stack-based language Michelson. There does appear to be ongoing work in the Tezos community to facilitate proofs about higher level representations of Tezos \contracts{} such as a step towards that with the intermediate language Albert~\cite{bernardo2020albert}.
Another formalization, Eth-isabelle~\cite{eth_isabelle} makes use of the same Lem model~\cite{lem_ethereum} as used in DeepSEA. While this theoretically means Eth-isabelle's capabilities are like those of DeepSEA, Eth-isabelle operates at a much lower level---typically reasoning about low-level EVM instructions which limits its expressiveness and usability, similarly to Mi-Cho-Coq.
The KEVM~\cite{hildenbrandt2018kevm} is a well validated formal semantics of the EVM which has been used along with the K-framework to enable formal verification of a range of real-world \contracts{}. The KEVM also relies upon an SMT solver to solve goals which would constrain its ability to prove some complex lemmas compared with a proof assistant such as Coq. To the best of our knowledge, the KEVM with the K framework is also only able to specify and prove reachability-related properties and so its specifications are less expressive than our approach allows.

\paragraph{Other Coq-based systems}
Scilla~\cite{sergey2019safer} is the \contract{} language of the Zilliqa blockchain. Their work on delivering a fully-functional proof system in Coq for Scilla has ceased (to the best of our knowledge). Nevertheless their work is helpful in demonstrating some key ideas about a \contract{} proof system.
ConCert~\cite{annenkov2020concert} is a \contract{} verification framework which enables the embedding of functional languages into Coq using a novel technique. Unlike DeepSEA, both approaches are not foundational.

\section{Conclusions and Future Work}

Using the DeepSEA system, users can verify blockchain contracts inside the Coq proof assistant. The generated model of the contract is a collection of Coq functions, which can be verified as easily as any other Coq program. At the same time, by using a verified compiler, the system can offer as strong foundational guarantees as if we were proving things directly about the compiled bytecode. We have applied DeepSEA to verify useful realistic programs, including a Uniswap-style market maker contract.

There are many directions for future work. First, although the language is already complete enough for useful contracts we would like to add more feautures, e.g. by completing the backend proofs about pointers and adding stack-allocated data to the surface language. Second, the verified compiler currently only deals with individual method calls, but we would like to extend the language semantics with a model of the full blockchain rather than leaving that model to the user as in Section~\ref{sec:crowdfunding}. Third, it will be interesting to explore more compiler optimizations, particularly because existing EVM compilers do not optimize very agressively. Most importantly, we will want to verify larger and more important blockchain applications, e.g. the many interacting contracts of a modern DeFi exchange.

\bibliography{references}

\end{document}